\documentclass{article}
\usepackage{graphicx}
\begin{document}
\rm
\let\sc=\sf

\begin{center}
\LARGE {\bf The Unique Eclipsing System KH 15D: New Photometric Data}

\vspace{1cm}
\Large {\bf O.Yu.\,Barsunova$^{1,3}$, V.P.\,Grinin$^{1,2,3}$,
S.G.\,Sergeev$^2$}
\end{center}
\large

1 - Central Astronomical Observatory Pulkovo, St. Petersburg, Russia,

2 - Crimean Astrophysical Observatory, Crimea, Nauchny, Ukraine,

3 - The Sobolev Astronomical Institute, St. Petersburg University, Russia\\

\begin{center}
\LARGE{}
\end{center}

\begin{abstract}
We present results of the photometric observations of the young
eclipsing binary KH 15D obtained during the two observational
seasons of 2002-2004 years. A comparison of our data with those by
Hamilton et al. (2001) and Herbst et al. (2002) reveals an
existence of the long-term photometrical brightness trend: during
5 years a system brightness in the I band decreased by about one
stellar magnitude. It is also shown that a systematic change of
the eclipse parameters found by Herbst et al. (2002) is continuing up
to now. The shape of the light curve did not change essentially
and is characterized with a small brightening in the central part
of the eclipses. The results obtained are discussed in the context
of the current models of KH 15D.
\end{abstract}

\section{Introduction}
An object KH 15D (Sp = K7; V = 18 - 21.5) has been discovered in
1997 by Kearns \& Herbst (1998) during a photometric study of the
young cluster NGC 2264. According to Hamilton et al. (2001) this
is a weak T Tauri star (WTTS). Its brightness fails by
approximately of 3.5 stellar magnitude with a periodicity of 48.35
d and in the failed state the object spends about one-third of the
period. Such eclipse parameters have no analog among the eclipsing
binaries that permits us to refer KH 15D to the unique
astrophysical objects.  During the last year it has been published
a series of important results which essentially completed an
information about this object. First of all, the fact of binarity
has been established: using such telescopes as Keck, Magellan II
and some others, Johnson et al. (2004) revealed a periodical
changes of the radial velocity of the object with an amplitude of
10.7 km/s and estimated the most probable ranges of the system
mass function ($0.125 \le f_M/\sin^3 i \le 0.5M_\odot$) and an
eccentricity $(0.68 \le e \le 0.80)$. Using the Harvard collection
of the photographic observations of NGC 2264, Winn et al. (2004)
found that at the beginning of 20th century the light curve of
this object strongly differed from that at the present time: an
effective duration of eclipses was essentially shorter or they
were not observed at all.

From the spectra in the region of 2 $\mu$m Tokunaga et al. (2004)
and Deming et al.(2004) found a bipolar outflow in the $H_2$
molecular lines. The latter authors concluded that the outflow
axis is noticeably inclined to the line-of-sight. Spectral
observations of Hamilton et al.(2003) at the bright state and
during eclipses showed that an equivalent width of the $H_2$ line
as well as those of forbidden lines [OI] 6300/6363 increases
synchronously with the decrease of the system brightness at the
moments of eclipses. The linear polarization has a similar
behavior (Agol et al. 2003). Such a behavior is typical for
objects whose source of the continuous radiation completely
screened from the observer at the moment of the eclipse, while a
more extended region forming a scattered polarized and the
emission line radiation is not screened or screened only partly. A
similar coronagraphic effect is observed in a clearest form in UX
Ori type stars and caused by episodic obscurations of the stars by
circumstellar gas and dust clouds (see the review by Grinin 2000,
Rodgers 2002 and the literature cited there).

To explain the unusual eclipses of KH 15D several models have been
proposed; their review is given by Winn et al. (2003) (see also
below). They give different predictions of the photometric
behavior of the system, that is why it is interesting to
investigate the photometry of this binary for a long time. With
this purpose, in 2002 we began the photometric observations of KH
15D. In this paper we present results obtained during the two
observational seasons of 2002-2004 years.

\section{Observations}
Observations were fulfilled at the AZT-8 0.7 m telescope in the
Crimean Astrophysical Observatory (CrAO). The detector was the CCD
camera AP7p (512 $\times$ 512 pixels, a pixel size was 24 $\times$
24$\mu$m ) established in the primary telescope focus. The
observations were carried out in the tree bands $v, r, i$ and
reduced into the Johnson-Cousins photometric system $V, R_c, I_c$.
Since the object of interest is in the young cluster NGC 2264,
practically all stars in its vicinity demonstrate brightness
fluctuations with the different amplitude. We have chosen four
stars for the photometric reference set which gave a minimal mean
square amplitude of the variation relatively to each other during
the time range considered. Table 1 contents numbers of these stars
according to the catalogue by Flaccomio et al. (1999) as well as
stellar magnitudes in the $I_c$  band. The rest columns give mean
$I_c$ values and their dispersions which we obtained in our
observations for two subsequent seasons. It is seen that these
values agree well with those published by Flaccomio et al. (1999).
Using all the four comparison stars permits us to degrade an
influence of the brightness fluctuation for each of them and
assure an accuracy of the photometry in the $I_c$ band not less
than $0.^m03$.

\begin{table}
\caption[]{Comparison stars} \label{ ta: one}
\begin{flushleft}
\begin{tabular}[center]{cccccc}
\hline\noalign{\smallskip}
N&$I_{Fl}$ & $I_1$&$\sigma$&$I_2$&$\sigma$  \\
\noalign{\smallskip}
\hline\noalign{\smallskip}
\noalign{\smallskip}
281&13.60&13.57&0.03&13.62&0.04 \\
328&14.53&14.48&0.04&14.47&0.04 \\
353&13.63&13.74&0.03&13.70&0.04 \\
422&13.66&13.62&0.05&13.62&0.05 \\
\noalign{\smallskip}
\hline\noalign{\smallskip}
\hline
\end{tabular}
\end{flushleft}
{\small Note: meanings of $I_{Fl}$ are taken from Flaccomio et al.
(1999).}
\end{table}

\subsection{Accounting the sky background}
One can see from Fig. 1 that near KH 15D there is a bright star
($V \approx 6$) whose scattered light is characterized with a
strong gradient at the location of the object of interest. This is
a serious obstacle in the correct accounting the sky background,
especially in the brightness minima. Therefore, a standard
procedure of the aperture photometry is not valid in this case
because the brightness of the sky background under the object can
noticeably differs from that we determine using the ring around
it.

\begin{figure*}
\vspace{-5.5cm}
\centering
\includegraphics[width=14cm]{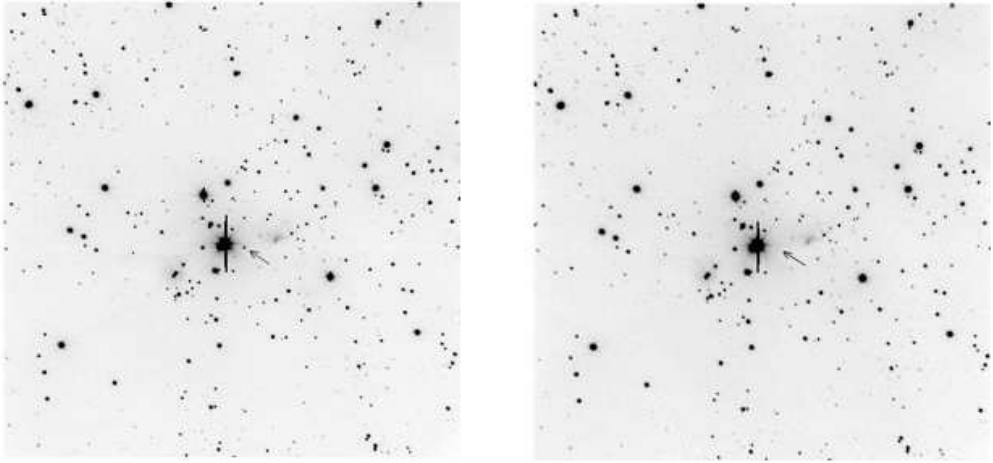}
\vspace{-6.5cm} \caption{ Image of KH 15D in the $I_c$ band in the
bright state (left) and during the eclipse (right).}
\end{figure*}

This problem can be solved by several ways from which we chose the
photometry with the instrumental profile PSF (Point Spread
Function). Let us suppose, that in the close vicinity of KH 15D
the scattered light is described with some plane defined by three
free parameters, say a central intensity and gradients on $X$ and
$Y$. Also, let us have the PSF for the image. Then, selecting the
position and intensity of the PSF as well as the parameters of the
plane yields the best approach to the brightness distribution
observed in this small region and hence, permits to find the
brightness of the object itself using the fitted PSF intensity.
The task is reduced to the multidimensional optimization of the
objective function. For this we developed and carefully tested a
special algorithm. In practice we varied only the position and
intensity of the PSF; the plane parameters were shaped due to the
plane approximation of the residual after a substraction of the
PSF. The size of the locality was $5 \times 5$ pixels or $8.8
\times 8.8$ arcseconds. The PSF itself was automatically
determined beforehand for each image by selection of the PSF for
the single stars. After averaging the PSF over all these stars we
condemned those of them which did not correspond to the mean PSF
due to the criterium of the mean square dispersion.

Comparison of the two methods (an aperture photometry and PSF)
showed that an essential discrepancy between them took place only
in the brightness minimum of KH 15D, and the first method often
gives formally negative brightness of the object thus resulting
the brightness of the sky background. In the bright state both
tools lead to close results with the mean square dispersion of
0.03 of the stellar magnitude.

An accuracy of the photometry was determined by with the
Monte-Carlo method i.e., frequentative introduction of the Gauss
noise into the image. The error in each pixel was found from the
quanta statistically accumulated, taking into account the pickup
noises and operations made with the certain image during its
treating. To this error we added (in the quadrature) the error of
the stellar magnitude caused by the method inaccuracy mentioned
above (0.03 of the stellar magnitude). The resulting accuracy of
the KH 15D photometry in its bright state was conditioned mainly
by the brightness fluctuations of the comparison stars and amounts
to $0.^{m}$03; in the weak state it is not worse than $0.^{m}$2.

\section{Results}
Figs. 2 and 3 demonstrate results of the KH 15D photometry in the
$I_c$ band based on our data and also on the data by Hamilton et
al. (2001) and Herbst et al. (2002). The photometric observations
in these two papers are given in the Cousins system like ours. An
accuracy of the measurements was about $0.^{m}01 - 0.^{m}02$ in
the bright state and $0.^{m}1 - 0.^{m}2$ in the weak state.

We see the photometric trend: the brightness of KH 15D
monotonously decreases with time. We discovered this trend
analyzing our observations: as it is seen from Fig. 2, during the
observational season of 2002-2003 yrs and partially in the next
season the object's brightness was droningly declining out of
eclipses. Supplementing the data kindly provided by W. Herbst to
our data we ascertained that this trend accorded well to the
brightness behavior of the object in the previous observational
season 2001-2002 yrs (as we can see from Fig. 2, it is also in
evidence in the behavior of the bright state of the object in
season of 2001-2002 yrs, when observations of KH 15D were carried
out by several groups (Herbst et al. (2002)) and were closest in
time.

\begin{figure*}
\vspace{-1.5cm}
\centering
\includegraphics[angle=-90,width=12cm]{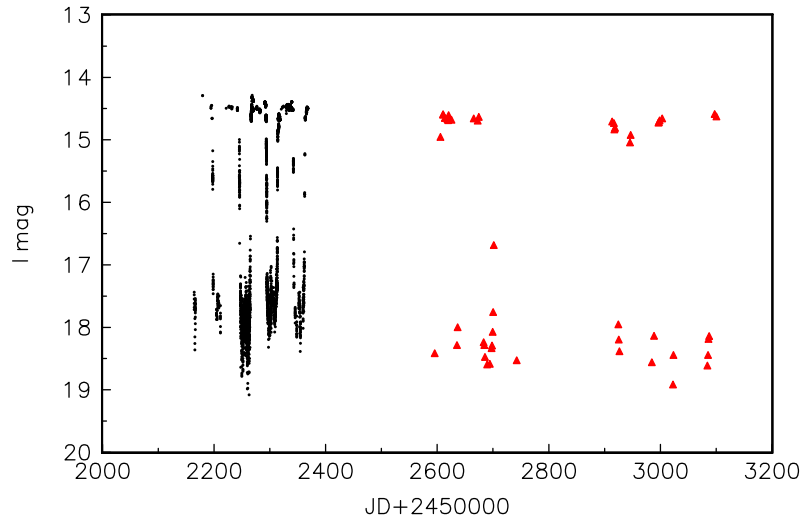}
\caption{ Light curve of KH 15D in the Cousins I band based on the
data of Herbst et al. (2002) (dots) and ours (triangles).}
\end{figure*}

Further supplementing these data by those obtained in 1999-2000
yrs by Hamilton et al. (2001) yields (see Fig. 3a) that the
earlier observations of this objects also reveal a brightness
behavior of analogous tendency.

As one can see from Fig. 3a, the brightness of KH 15D out of
eclipses has diminished by approximately $0.^{m}$7 for the two
seasons of 1999-2001 while during next three observational seasons
(2001-2004) the object has failed by about $0.^{m}$3. This implies
a retardation of the photometric trend and some points on the
light curve obtained at the end of the observational season of
2003-2004 yrs (Fig. 2) indicate to its possible completion and
beginning of the brightness increase of KH 15D. Such a light
behavior (if it will be confirmed by farther observations) can
serve an important clue for understanding a mechanism of the
eclipses of this object (see below).

On Fig. 3 we give convolutions of the light curve of the object
with the period of 48.35 days for the three time ranges mentioned
above. The brightness trend of KH 15D is distinctly seen. One can
see also that the increase of the eclipse duration noticed firstly
by Hamilton et al. (2001) occurs up to date. From Fig. 3b, where
the light curves are given after the trend removal we see that
diminution of the eclipse depth, observed by Hamilton et al.
(2001) and Herbst et al. (2002), has decelerated  or fully stopped
in 2002-2004. But the shape of the light curve in total did not
change. In the central part of the eclipse a weak brightening
takes place as before apparently being a special feature of the
light curves of the object (Hamilton et al. (2001),Herbst et al.
(2002)). Analyzing our observations in the $V$ and $R_c$ bands we
proved the conclusion of Hamilton et al. (2001) about a neutral
character of the eclipses of KH 15D.

\begin{figure*}
\vspace{-1.5cm}
\hspace{-3.5cm}
\includegraphics[angle=-90,width=23cm]{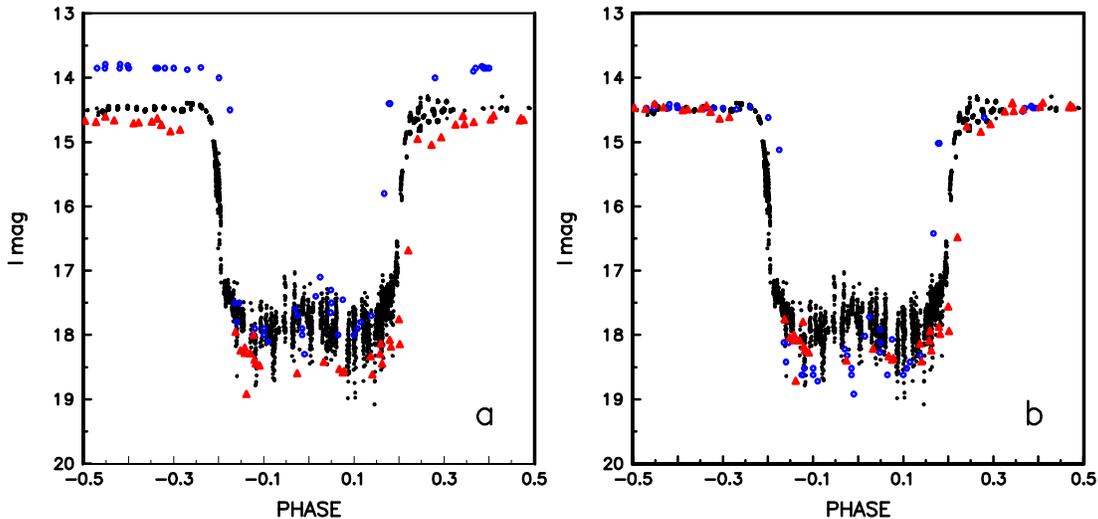}
\vspace{-6cm} \caption{ a) Light curve of KH 15D â in the $I_c$
band convolved with a period of 48.36 d; b) the same as a) after
accounting the trend.  Circles denote data of Hamilton et al.
(2001) for 1999-2000 yrs, dots are the data of Herbst et al.
(2002) for 2001-2002 yrs, triangles are our data for 2002-2004
yrs.}
\end{figure*}

As we can see from Fig. 3, the radiation flux from the object
fluctuates at the moment of eclipse with an amplitude of about one
stellar magnitude. It follows from the light curve obtained by
Herbst et al. (2002) in 2001-2002 yrs when such fluctuations were
observed rather often due to the dense monitoring. We also fixed
one episode when the flux strongly deviated from the average light
curve during the eclipse. Since in the light minima the radiation
scattered by the circumstellar dust mainly contributes to the flux
observed (a growth of the linear polarization in the minima
justifies this (Agol et al. (2003)), the special behavior of the
object's brightness during the eclipse pointed out above testifies
that parameters of the scattered radiation of KH 15D are undergone
strong fluctuations.

\section{Discussion}
The new and the most important result of our analysis is a
conclusion about an existence of the trend of the KH 15D
brightness: in total, for the last 5 years the object faded by
about one stellar magnitude. When we have already been obtained
this result, Johnson and Winn (2003) published results of their
analysis of the photographic observations of NGC 2264 cluster
carried out at the observatory Asiago in 1967 - 1982. They found
that at that epoch KH 15D in the $I_c$ band was brighter by $\sim
0.^{m}9$ of stellar magnitude compared to the data of Herbst et
al. (2002) in 2001-2002. Taking into consideration that the
brightness trend we discovered is comparable in amplitude with
that determined by Johnson and Winn (2003) but is reached during
the essentially shorter time, one can suppose that it has a cyclic
character and, apparently, duration of the cycle significantly
less then 30 years. The deceleration of the trend noted above
indirectly confirms this suggestion.

How to explain this trend of the brightness in the framework of
the models elaborated for this eclipsing system?

Obviously, one of the possible reasons for the slow changes of the
object's brightness out of eclipse can be a change of the
luminosity of the primary component of the binary system. Although
we cannot to specify a concrete mechanism caused such variations,
nevertheless, one cannot reject completely such an explanation.
Another possible cause is an additional, slowly varied with time
extinction of the radiation from the primary on the way to an
observer.

Based on the observed variations of the eclipse parameters, Winn
et al. (2004) and Chiang and Murray-Clay (2004) proposed the model
of KH 15D in which the binary system consists of two components
with approximately the same parameters and is surrounded with and
opaque gas and dust ring or disk (farther for brevity we shall
call it circumbinary (CB) disk. It is assumed that the system is
seen nearly edge-on while the CB disk is inclined to the
line-of-sight on about 20$^{\circ}$ and slowly precesses
relatively to the mass center of the system. Eclipses occur when
one or both companions turn up in "the shadow" of the CB disk
during their orbital motion. In order to explain the sharp ingress
and egress from the minimum, the authors claim a rather steep
internal boundary of the CB disk.

The changes of the parameters of the eclipse in this model is
attributed to the CB disk precession. It is suggested that in the
modern epoch one of the component has already completely closed by
the CB disk and only near the periastron its radiation slightly
shines trough the disk (that explains a small rise of the
brightness in the central part of the eclipse). Eclipses befall
when the secondary sets behind the opaque boundary of the disk in
its orbital motion.

Grinin and Tambovtseva (2002) suggested that the disk wind of the
secondary, opaque due to presence of the dust can cause the long
lasting eclipses. They supposed that the plane of the binary
(coinciding with the plane of the CB disk) is inclined to the
line-of-sight. Calculations show (Grinin and Tambovtseva (2002);
Grinin et al. (2004)) that if the dust in the wind has parameters
typical for the circumstellar disks and well mixed with the gas in
a proportion 1:100 as in the interstellar medium, then the
amplitude of the brightness decrease observed in KH 15D is
obtained at the mass loss rate of the disk wind (in the low
velocity component) of about $3 \cdot 10^{-8} \dot{M_{\odot}}$
yr$^{-1}$. A change of the eclipse parameters in this model takes
place as a result of the motion of the apsis line of the
secondary's orbit (Grinin et al. (2004)) but can be also caused by
the wind instabilities. Brightening observed in the central part
of an eclipse is due to low density of the matter in the interiors
of the disk wind cone.

Since CB disks in the binaries with eccentric orbits are described
with a global asymmetry (Artymowicz and Lubow (2000)), then, under
a small inclination to the line-of-sight, in such systems along
with eclipses caused by the disk wind of the secondary one can
also observe essentially slower brightness variations (similar to
that we found in KH 15D) attributed to changes of extinction on
the line-of-sight due to the CB disk precession.

\section{Conclusion}
Thus, we showed that in the modern epoch there is trend of the
brightness in the light variations of KH 15D: during 5 years this
object has failed by about one stellar magnitude. Analyzing the
bright state in the last observational season we suspect that this
trend slowed down and some cue on the opposite tendency in the
brightness behavior appeared (Fig. 2). If this tendency will be
confirmed by future observations it will be possible to certify
that "the end of light" predicted by the models of Winn et al.
(2004) and Chiang and Murray-Clay (2004), when both components of
the system will be completely closed from an observer by the CB
disk (according to estimates of Winn et al. it will be in 2012)
will not take place.

The results of the analysis given above show that in order to
clarify the essence of unusual eclipses of KH 15D it is necessary
to have an information about light curves not only at the moment
of eclipses (which is mainly interested by observers) but also out
of them. It would be highly interesting also to carry out a
photometric monitoring of this unique object in the near infrared
region of spectrum.\\

\textbf{Acknowledgements}\\
We thank V. Doroshenko and E. Sergeeva for the assistance in
observations, Bill Herbst for providing us with the data of the
photometric observations of KH 15D and also V.M. Larionov, the
referee of the paper for his useful comments. The work is
supported by the grant of Presidium of RAS "Non-stationary
phenomena in Astronomy" and the grant INTAS N 03-51-6311.

\begin{center}
{\bf References}
\end{center}
\large
E. Agol, A. Barth, S. Wolf, D. Charbonneau, 2003, astro-ph 0309309\\
P. Artymowicz, S.H. Lubow, 1996, Astrophys. J. {\bf 467} L77 \\
E.I. Chiang, R.A. Murray-Clay, 2004, astro-ph 0312515\\
D. Deming, D. Charbonneau, J. Harrington, 2004, Astrophys. J.,
{\bf 601}, L87\\
E. Flaccomio, G. Micela, S. Sciortino, F. Favata, C. Corbally, A.
Tomaney, 1999, A\&A, {\bf 345}, 521\\
V.P. Grinin, 2000, in "Disk, Planetesimal and Planets", Eds. by
F.Garzon, et al. ASPC, {\bf 219}, p. 216 \\
V.P.Grinin, L.V. Tambovtseva, 2002, Astron. Letters, {\bf 28},
601\\
V.P.Grinin, L.V. Tambovtseva, N.Ya. Sotnikova, 2004, Astron.
Letters. \textbf{30}, 694\\
C.M. Hamilton, W. Herbst, C. Shih, A.J. Ferro, 2001, ApJ
\textbf{554}, L201 \\
C.M. Hamilton, W. Herbst, R. Mundt, C.A.L. Bailer-Jones, C.M.
Johns-Krull, 2003, Astrophys. J., {\bf 591}, L45 \\
W. Herbst, C.M. Hamilton, F.J. Vrba et al. 2002, PASP \textbf{114}, 1167 \\
J.A. Johnson, J.N. Winn, 2003, astro-ph 0312428\\
J.A. Johnson, G.W. Marcy, C.M. Hamilton, et al. 2004, astro-ph 0403099\\
K.E. Kearns, N.L. Eaton, W. Herbst, C.J. Mazzurco 1997, AJ \textbf{114}, 1098\\
B. Rodgers, D. Wooden, V.P. Grinin, D. Shakhovskoj, A. Natta,
2002, Astrophys. J., {\bf 564}, 405\\
A.T. Tokunaga, S. Dahm, W. G\"assler et al. 2004, astro-ph 0401177\\
J.N. Winn, P.M. Garnavich, K.Z. Stanek, D.D. Sasselov, 2003, ApJ
{\bf 593}, L121\\
J.N. Winn, M.J. Holman, J.A. Johnson et al. 2004, astro-ph 0312458\\

\end{document}